\documentclass[final]{pasj00}
\usepackage{graphicx}

\SetRunningHead{T. Akahori and K. Yoshikawa}{Non-Equilibrium
  Ionization State and Two-Temperature Structure in Abell 399/401}

\title{Non-Equilibrium Ionization State and Two-Temperature Structure\\
in the Linked Region of Abell 399/401}

\author{Takuya \textsc{Akahori} and Kohji \textsc{Yoshikawa}}
\affil{Center for Computational Sciences, University of Tsukuba, 1-1-1, Tennodai, Tsukuba, Ibaraki 305-8577}
\email{akataku@ccs.tsukuba.ac.jp, kohji@ccs.tsukuba.ac.jp}


\Received{2008 May 16}
\Accepted{2008 June 2}
\Published{}

\begin{document}

\maketitle

\begin{abstract}
  We investigate a non-equilibrium ionization state and
  two-temperature structure of the intracluster medium in the linked
  region of Abell 399/401, using a series of N-body + SPH simulations,
  and find that there exist significant shock layers at the edge of
  the linked region, and that the ionization state of iron departs
  from the ionization equilibrium state at the shock layers and around
  the center of the linked region. As for the two-temperature
  structure, an obvious difference of temperature between electrons
  and ions is found in the edge of the linked regions. K$\alpha$ line
  emissions of Fe\,\textsc{xxiv} and Fe\,\textsc{xxv} are not severely
  affected by the deviation from the ionization equilibrium state
  around the center of the linked region, suggesting that the
  detection of relatively high metallicity in this area cannot be
  ascribed to the non-equilibrium ionization state of the intracluster
  medium. On the other hand, the K$\alpha$ emissions are significantly
  deviated from the equilibrium values at the shock layers, and the
  intensity ratio of K$\alpha$ lines between Fe\,\textsc{xxiv--xxv}
  and Fe\,\textsc{xxvi} is found to be significantly altered from that
  in the ionization equilibrium state.
\end{abstract}

\section{Introduction}
According to a standard scenario of hierarchical structure formation
in the universe, galaxy clusters are formed through successive merging
of galaxies, galaxy groups, and clusters.  About a half of observed
clusters have irregular X-ray morphology (e.g., \cite{AM05} references
therein), and a part of such irregularities thought to be caused by
such merging events.

Abell 399/401 is well-known as merging clusters on an early stage of
the merging. 
\citet{Fujita96} and \citet{SP04} found that the intracluster medium
(ICM) at the linked region of the two clusters is compressed based on
the ASCA and XMM-Newton observations, respectively.
Recently, \citet{Fujita08} reported the detection of Fe K emission
lines near the center of the linked region based on the Suzaku XIS
observation, and the metallicity in that regions is estimated to be
0.2 times the solar metallicity, which seems relatively high compared
with theoretical predictions based on numerical simulations
(e.g., \cite{Tornatore2007}).

The estimation of physical properties of ICM in X-ray observations is
usually based on the assumptions that ICM is in the ionization
equilibrium state and that electrons and ions share the same thermal
temperature. Such assumptions can be justified around central regions
of galaxy clusters by the fact that ICM density is high enough to
quickly achieve ionization equilibrium and thermal equilibration
between electrons and ions.  As for the ionization equilibrium, the
timescale required to reach collisional ionization equilibrium for an
ionizing plasma, is estimated as $n_{\rm e}t\gtrsim 10^{12}~{\rm
  cm^{-3}~s}$ (see e.g., \cite{Masai84}), where $n_{\rm e}$ is the
number density of electrons.  Therefore, for $n_{\rm e}\sim
10^{-4}~{\rm cm^{-3}}$ in the linked region (\cite{SP04};
\cite{Fujita08}), $t \sim {\rm Gyr}$ is comparable to or longer than
the merger timescale, so that the ionization equilibrium is no longer
a reasonable assumption.  Actually, in the warm-hot intergalactic
medium (WHIM), where the density is much lower, the deviation from the
ionization equilibrium have been pointed out by \citet{YS06}. In
addition, the thermal equilibration in merging clusters has been
studied by \citet{Takizawa99} based on N-body/SPH simulations, and the
two-temperature structure of ICM, i.e. the difference in temperature
between electrons and ions, is reported especially at low-density
regions. The two-temperature structure of WHIM has been also suggested
by \citet{Yoshida05}.

Therefore, there exists sufficient reasons to suspect that these
assumptions are not valid in the linked region of Abell 399/401. In
this paper, we investigate an ionization state and temperature
structure of ICM in the linked region by relaxing the assumptions of
ionization equilibrium and thermal equipartition between electrons and
ions, and verify to what extent such a non-equilibrium state affect
the interpretations of the observational data.

\section{Model and Calculation}

We carry out N-body/SPH simulations of two merging galaxy clusters, in
which a non-equilibrium ionization state and two-temperature structure
of ICM are both taken into account. Radiative cooling and electron
heat conduction are both ignored in these simulations.

The time evolution of two-temperature structure is followed with the
same way as \citet{Takizawa99}. We assume that the electrons and ions
always reach Maxwellian distributions with temperatures, $T_{\rm e}$
and $T_{\rm i}$, respectively, and the two temperatures are equalized
through Coulomb scattering on a timescale of
\begin{equation}
t_{\rm ei}=2\times 10^8~{\rm yr}
\frac{(T_{\rm e}/10^8~{\rm K})^{3/2}}{(n_{\rm i}/10^{-3}~{\rm cm^{-3}})}
\cdot\left(\frac{40}{\ln\Lambda}\right),
\end{equation}
where $n_{\rm i}$ is the number density of ions, and $\ln \Lambda$ is
the Coulomb logarithm. By introducing the dimensionless temperatures
of electrons and ions, $\tilde{T}_{\rm e}\equiv T_{\rm e}/T$ and
$\tilde{T}_{\rm i}\equiv T_{\rm i}/T$, respectively, normalized by the
mean temperature of electrons and ions, $T\equiv (n_{\rm e}T_{\rm
  e}+n_{\rm i}T_{\rm i})/(n_{\rm e}+n_{\rm i})$, the evolution of
two-temperature structure is described by
\begin{equation}
\frac{d\tilde{T}_{\rm e}}{dt}=
\frac{\tilde{T}_{\rm i}-\tilde{T}_{\rm e}}{t_{\rm ei}}
-\frac{\tilde{T}_{\rm e}}{u}Q_{\rm sh},
\end{equation}
where $u$ and $Q_{\rm sh}$ are the specific thermal energy and the
shock heating rate per unit mass, respectively.  The first term of the
r.h.s. denotes the thermal relaxation rate between ions and electrons.
Note that recent studies on supernova remnants showed that $T_{\rm e}/T_{\rm i}\sim 1$
around the shock for relatively small Mach numbers (\cite{Ghavamian07});
electrons may be heated by some plasma waves. However, for
a cluster merger, the Mach number of the shock is `very' small
($<5$). In this case, waves may not be generated and electrons may not
be heated. In this letter, we ignore the effect of electron heating by
plasma waves.

The time evolution of ionization fractions of ions is computed by
solving
\begin{eqnarray}
\frac{df_j}{dt}&=&\sum_{k=1}^{j-1}S_{j-k,k}f_k-
\sum_{i=j+1}^{Z+1}S_{i-j,j}f_j\nonumber\\
&-&\alpha_jf_j+\alpha_{j+1}f_{j+1},
\end{eqnarray}
where $j$ is the index of a particular ionization stage considered,
$Z$ the atomic number, $f_j$ the ionization fraction of an ion $j$,
$S_{i,j}$ the ionization rate of an ion $j$ with the ejection of $i$
electrons, and $\alpha_j$ is the recombination rate of an ion $j$.
Ionization processes include collisional, Auger, charge-transfer, and
photo-ionizations, and recombination processes are composed of
radiative and dielectronic recombinations.  Ionization and
recombination rates are calculated by utilizing the SPEX ver 1.10
software package\footnote{http://www.sron.nl/divisions/hea/spex/}.
Actual calculations are carried out in essentially the same way as
\citet{YS06}, except that the reaction rates are computed using the
electron temperature, $T_{\rm e}$, rather than the mean temperature, $T$,
in order to incorporate the effect of two-temperature structure.
We solve the time evolution of each ionization fraction of H, He,
C, N, O, Ne, Mg, Si, S, and Fe, but we focus only on iron.

To reproduce the situation of Abell 399/401, the initial condition of
the two clusters is set up as follows. We consider a head-on merger
(\cite{OH94}),
and assume that their collision axis is perpendicular to the line of sight.
Each cluster has the same shape for simplicity since the differences
of the two clusters in shape and size are slight (\cite{SP04}).  We
adopt the King profile for the initial dark matter distribution, in
which the radial velocity dispersion, $\sigma_{*r}$, satisfies
$\sigma_{*r}^2=k_{\rm B}T_{\rm vir}/\mu m$, where $k_{\rm B}$ is the
Boltzmann constant and $T_{\rm vir}$ is the virial temperature (the
virial radius is $r_{\rm vir}=3.0$~Mpc and the cluster mass is
$M(r_{\rm vir})\simeq 1.7\times 10^{15}~\MO$).  For the initial ICM
distribution, we first assume the temperature profile,
$T(r)=(T_{\rm vir}/\beta)\exp(-r/r_{\rm vir})$,
with $\beta =0.6$, and the density profile is obtained by assuming
hydrostatic equilibrium.
The above parameters are chosen so that X-ray
surface brightness satisfies the $\beta$-model best-fit \citep{SP04},
and that the spectroscopic-like temperature, $T_{\rm sl}$
\citep{Mazzotta04}, reproduces the observed temperature at the
unperturbed regions by \citet{Mark98}.
The number of particles of each galaxy cluster is set to a half
million each for dark matter and ICM.  The initial relative velocity
of the two clusters is set to $1050~{\rm km/s}$ to reproduce the
observed temperature of the linked region in terms of $T_{\rm sl}$,
and the initial separation is $\sim 7$~Mpc. As for the metallicity, a
spatially uniform metallicity of 0.2 times the solar abundance is
assumed. It should be noticed that the metallicity, $Z$, is very
insensitive to the resulting ionization state of ions as long as $Z\ll
1$.  We also assume $\tilde{T}_{\rm e}=\tilde{T}_{\rm i}=1$ and an
ionization equilibrium state at the start of the simulation.

\section{Result}

\begin{figure*}[tb]
\begin{center}
\FigureFile(135mm,75mm){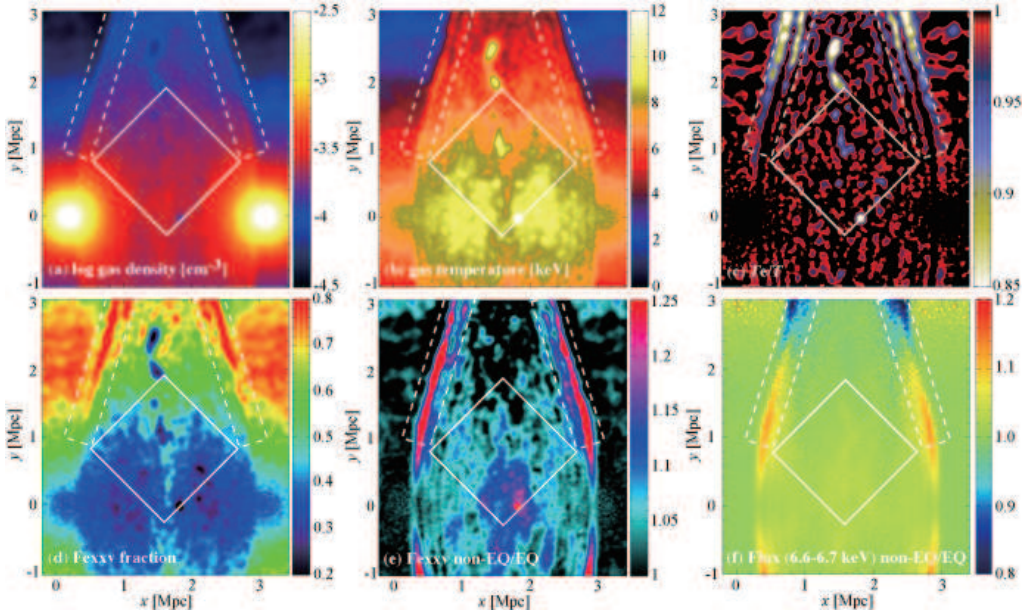}
\end{center}
\caption{Maps of (a) the ICM density in units of $\log~{\rm cm^{-3}}$,
  (b) the mean temperature in keV, (c) the ratio of the electron
  temperature relative to the mean temperature, (d) the ionization
  fraction of Fe\,\textsc{xxv}, (e) the ratio of the Fe\,\textsc{xxv}
  fraction relative to that in the ionization equilibrium state, and
  (f) the ratio of the line intensity in 6.6--6.7~keV band integrated
  along the line of sight relative to that in the equilibrium state.
  The white-solid squares and white-dashed rectangles
  indicate the regions corresponding to the Suzaku observation by
  \citet{Fujita08} and the shock layers, respectively.  }
\end{figure*}

In the rest of this paper, we present the results of a snapshot of the
simulation in which the separation between the centers of the two
clusters is $\sim 3$~Mpc so that their configuration is consistent
with that of Abell 399/401. Figure 1a--1e show physical properties
on a cross section along the collision axis of the two clusters.
The white-solid square with 1.5~Mpc on a side shows the region roughly
corresponding to the observed area by \citet{Fujita08}.  In this
region, the local ICM density, $n~(= n_{\rm e}+n_{\rm i})\sim 3\times
10^{-4}~{\rm cm^{-3}}$ (figure 1a), and $T_{\rm sl}=6.48~{\rm keV}$
are in good agreement with the observations (\cite{SP04}; \cite{Fujita08}).

We find that the shock heating rate is typically a few percent of the
adiabatic heating rate inside the linked region. This means that both
the electrons and ions are mainly heated by the adiabatic compression
in almost the same rate with no significant shocks there. Accordingly,
as can be seen in figure 1c, the electron temperature is only a few
percent lower than the mean temperature in this region. On the other
hand, at the edge of the linked region
(the white-dashed rectangles), there exist significant shock
layers with mach number of 1.5--2. In these layers, we can see that
the electron temperature is typically 10--20~\% lower than the mean
temperature, and even lower toward outer regions.

Figure 1d and 1e show the ionization fraction of Fe\,\textsc{xxv} and
the ratio of the fraction relative to that in the ionization
equilibrium state, respectively.  Here, the ionization equilibrium
state is calculated assuming that $T_{\rm e} = T_{\rm i}$.  In the
center of the linked region, the Fe\,\textsc{xxv} fraction is
typically 30--60~\%, the largest among other ionization states, and is
10--20~\% larger than that in the equilibrium state. In the shock
layers at the edge of the linked region, the Fe\,\textsc{xxv} fraction
is nearly 80 \% and 30--40 \% larger than that in the ionization
equilibrium state.

The excess of the Fe\,\textsc{xxv} fraction can be understood as
follows. In the linked region, the electron temperature is increasing
from $\sim 3$~keV to $\sim 8$~keV according to the adiabatic
compression and/or the shock heating, so that the Fe\,\textsc{xxv}
fraction is decreasing in time because an Fe\,\textsc{xxv} fraction
peaks at $\sim 3$~keV and decreases as the temperature increases in
the ionization equilibrium state. However, the ionization of
Fe\,\textsc{xxv} to higher ionization states is not quick enough to
catch up with the ionization equilibrium state, leaving the
Fe\,\textsc{xxv} fraction larger than that in the equilibrium state.

Deviations from the ionization equilibrium state can be seen also for
Fe\,\textsc{xxiv}.  Its fraction is $\simeq 2$--3~\%, and $\sim
15$--20~\% larger than that in the equilibrium state. Due to the
excess of Fe\,\textsc{xxiv} and Fe\,\textsc{xxv} fractions, the
fractions of Fe\,\textsc{xxvi} and Fe\,\textsc{xxvii} are $\sim 5$~\%
and $\sim 15$--20~\% smaller than those in the equilibrium state,
respectively.

As a result of the deviation from the ionization equilibrium, the
intensities of iron line emissions are altered to some extent.
We calculate X-ray spectrum from the simulation using the 
SPEX package, and find that
K$\alpha$ lines of Fe\,\textsc{xxiv} and Fe\,\textsc{xxv} in a
rest-frame energy band of 6.6--6.7 keV is larger than those in the
ionization equilibrium state, primarily because of the excess of
Fe\,\textsc{xxv} fraction. On the other hand, since the
Fe\,\textsc{xxvi} fraction is smaller than that in the equilibrium
state, K$\alpha$ lines of Fe\,\textsc{xxvi} in 6.9--7.0 keV band in
rest frame are dimmer.  It should be noted that the iron line emission
detected in \citet{Fujita08} corresponds to the K$\alpha$ lines of
Fe\,\textsc{xxiv} and Fe\,\textsc{xxv} in the 6.6--6.7 keV energy
band.  Figure~1f depicts the ratio of the intensity in 6.6--6.7 keV
band projected along the line of sight relative to that in the
ionization equilibrium state, and it can be seen that the intensity
changes are significant (typically $\sim 15$ \%) at the shock layers,
while the intensity is only a few percent enhanced around the area of
the Suzaku observation by \citet{Fujita08}, despite the excess of the
Fe\,\textsc{xxv} fraction by 10--20 \% (figure 1e). This is because
the deviation from the ionization equilibrium is significant only at
the center of the linked region, and its effect is diluted in
integrating along the line of sight. This suggests that the iron line
emission detected by \citet{Fujita08} is not severely affected by the
deviation from the ionization equilibrium.  Here, let us define the
ratio of the X-ray intensity between the two energy bands as
\begin{equation}
  R=\frac{F(6.6-6.7~{\rm keV})}{F(6.9-7.0~{\rm keV})}.
\end{equation}
Figure~2 shows the map of the ratio $R/R_{\rm eq}$, where $R_{\rm eq}$
is the intensity ratio defined above but in the ionization equilibrium
state, and clearly indicates that the intensity ratio departs from the
one in the ionization equilibrium state within the shock layers. Note
that this intensity ratio is independent of the local abundance of
iron, but primarily depends on its ionization state. Therefore, such
excess of the intensity ratio, $R$, can be used as a stringent
observational tracer of the non-equilibrium ionization state.

\begin{figure}[tb]
\begin{center}
\FigureFile(55mm,35mm){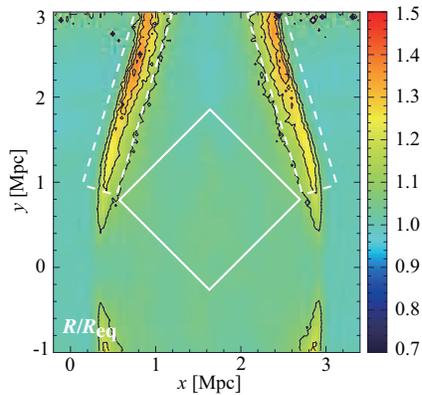}
\end{center}
\caption{The map of the ratio, $R/R_{\rm eq}$, in the same region as
  figure~1. The contours from outside to inside indicate $R/R_{\rm
    eq}=1.1$ to 1.4 separated by a difference of 0.1.}
\end{figure}

\section{Conclusion and Discussion}

The ionization state of iron and two-temperature structure in the
linked region of Abell 399/401 is investigated by using N-body + SPH
simulations by relaxing the assumptions of ionization equilibrium and
thermal equilibration between electrons and ions.

It is found that, around the center of the linked region, the
Fe\,\textsc{xxv} fraction is 10--20~\% larger than that in the
ionization equilibrium state, and the electron temperature is only a
few percent smaller than the mean temperature of electrons and ions,
because the ICM is mainly adiabatically heated inside the linked
region. On the other hand, we find that there exist shock layers at
the edge of the linked region, and that the Fe\,\textsc{xxv} fraction
is larger than that in the ionization equilibrium state by 30--40 \%,
and the electron temperature is typically 10--20~\% lower than the
mean temperature around these layers.

Our simulation indicates that the intensity of Fe~K$\alpha$ emission
lines is affected by such deviation from the ionization equilibrium
state of iron in the linked region. While the deviation from the
ionization equilibrium state is remarkable around the center and the
edge of the linked region, we find that the emission line intensity is
strongly affected preferentially at the edge of the linked region.

The area of the Suzaku XIS observation by \citet{Fujita08}, in which
fairly high metallicity is reported, is located at the central
portion of the linked region. 
Our results imply that X-ray emission in
this area is not strongly affected by the effects of a non-equilibrium
ionization state and two-temperature structure. Therefore, the fact
that high metallicity is detected in this area cannot be ascribed to
the non-equilibrium ionization state of the ICM, and must be explained
by other physical processes.

It is interesting to discuss the detectability of the non-equilibrium
ionization state and the two-temperature structure of ICM in Abell
399/401. According to our simulations, it is expected that shock
layers with a mach number of 1.5--2 are located at the edge of the
linked region, and that the ratio between K$\alpha$ emission lines of
Fe\,\textsc{xxiv}--\textsc{xxv} and Fe\,\textsc{xxvi} are
significantly different from that in the ionization equilibrium state.
Observationally, such an intensity ratio of iron emission lines could
be important clues for the detection of the deviation from the
ionization equilibrium in these layers. In principle, the indication
of the two-temperature structure can be also obtained by the
difference between electron and ion temperatures inferred from X-ray
thermal continuum and thermal width of emission lines, respectively.
Of course, the detections of the non-equilibrium ionization state and
the two-temperature structure are not very feasible with the current
observational facilities, but could be achieved by the X-ray
spectroscopy with a high energy resolution using X-ray calorimeters in
near future.

Finally, we should discuss several caveats on the assumptions adopted
in this work.  The slope of the density profile, i.e.  $\beta$, is a
sensitive parameter to the non-equilibrium ionization state and
two-temperature structure. If we adopt $\beta=0.5$, the resultant
departure from the ionization equilibrium is suppressed because the
ICM density at the outskirts is relatively denser and the
equilibration timescales become shorter. In addition, we assume that
the encounter is taking place on the plane of the sky. If it is not
the case, the effect of the non-equilibrium ionization state on the
observed X-ray intensity at the shock layers would be blurred to some
extent in integrating along the line of sight, depending on the
viewing angle.

\bigskip

The authors would like to thank an anonymous referee for useful comments and suggestions. 
This work is carried out with computational facilities at Center for
Computational Sciences in University of Tsukuba, and supported in part
by Grant-in-Aid for Specially Promoted Research (16002003) from MEXT
of Japan, and by Grant-in-Aid for Scientific Research (S) (20224002)
and for Young Scientists (Start-up) (19840008) from JSPS.

\end{document}